\documentclass[12pt]{article}
\usepackage{amsmath,amssymb}
\usepackage{color}
\renewcommand{\Large}{\large} % smaller headlines

\allowdisplaybreaks
\begin{document}
%%%%%%%%%%%%%%%%%%%%%%%%%%%%%%%%%%%%%%%%%%%%%%%%%%%%%%%%%%%%%
%
\begin{titlepage}
\bigskip

\begin{center}
%\vskip 6mm
%\vskip 3in
%\begin{document}
%\vskip 5pt
%\pagenumbering{roman}
\begin{center}
{\Large \bf Anisotropic expansion, second order hydrodynamics and
holographic dual}
\\[10mm]
\end{center}
%\vskip 8mm

\textbf{Priyanka Priyadarshini Pruseth
\footnote{
Present address: Department of Physics, Vesaj Patel College,
Sundargarh, India} and
Swapna Mahapatra}

\vskip 6mm

{\em Department of Physics, Utkal University,
Bhubaneswar 751004, India.}\\

{\tt pkpruseth@gmail.com}\;,\;\, {\tt swapna.mahapatra@gmail.com}
\end{center}

\vskip .2in
%%%%%%%%%%%%%%%%%%%%%%%%%%%%%%%%%%%%%%%%%%%%%%%%%%%%%%%%%%%%%%%%
\begin{center}
{\bf Abstract }
\end{center}
\begin{quotation}
\noindent
We consider Kasner space-time describing anisotropic
three dimensional expansion of RHIC and LHC fireball and study the
generalization of Bjorken's one dimensional expansion by taking into
account second order relativistic viscous hydrodynamics. Using time
dependent AdS/CFT correspondence, we study the late time behaviour
of the Bjorken flow. From the conditions of conformal invariance and
energy-momentum conservation, we obtain the
explicit expression for the energy density as a function of proper time
in terms of Kasner parameters. The proper time dependence of the
temperature and entropy have also been obtained in terms of Kasner parameters.
We consider Eddington-Finkelstein type coordinates and discuss the gravity
dual of the anisotropically expanding fluid in the
late time regime.

%%%%%%%%%%%%%%%%%%%%%%%%%%%%%%%%%%%%%%%%%%%%%%%%%%%%%%%%%%%%
\end{quotation}
\vfill
%%%%%%%%%%%%%%%%%%%%%%%%%%%%%%%%%%%%%%%%%%%%%%%%%%%%%%%%%%%%
%\today
%%%%%%%%%%%%%%%%%%%%%%%%%%%%%%%%%%%%%%%%%%%%%%%%%%%%%%%%%%
\end{titlepage}
%%%%%%%%%%%%%%%%%%%%%%%%%%%%%%%%%%%%%%%%%%%%%%%%%%%%%
%%%%%%%%%%%%%%%%%%%%%%%%%%%%%%%%%%%%%%%%%%%%%%%%%%%%%
\section{Introduction}
\label{sec:introduction}
\setcounter{equation}{0}
%%%%%%%%%%%%%%%%%%%%%%%%%%%%%%%%%%%%%%%%%%%%%%%%%%%%%% 
%\begin{Large}
%{\bf Introduction} \\
%\end{Large}

AdS/CFT correspondence has provided a very important tool for studying 
the strongly coupled dynamics in a class of superconformal field theories,
in particular, ${\cal N} = 4$  super Yang-Mills theory  
and the corresponding gravity dual description in AdS space-time 
\cite{Malda:1998, WittenGubser:1998}. 
One of the fundamental question in the field of high energy physics
is to understand the properties of matter at extreme density and
temperature in the first few microseconds after the big bang.
Such a state of matter is known as Quark-Gluon-Plasma (QGP) state where
the quarks and the gluons are in deconfined state. A lot of progress has 
been made in understanding the properties and various aspects of the 
evolution of strongly coupled QGP through the heavy ion collision 
experiments at RHIC \cite{Shuryak:2005, Kolb:2003KH} 
(see also \cite{Tannenbaum:2006}) as well as in LHC (see 
\cite{Florkowski:2017} for a comprehensive review). 
Hydrodynamics plays an important role after the system undergoes a 
rapid thermalization and local thermal equilibrium is reached. 
Since it is difficult to solve the strongly coupled QCD, the qualitative 
features of the hydrodynamics regime in the evolution of QGP has been 
studied by using the AdS/CFT duality. In the context of heavy ion 
collisions, AdS/CFT correspondence has led to very interesting results 
like computation of shear viscosity of 
finite temperature ${\cal N} = 4$ Supersymmetric Yang-Mills theory
\cite{Policastro:2001}, viscosity from gravity dual description involving 
black holes in AdS space  \cite{Kovtun:2005} etc.  
 
As we know, perfect local equilibrium  system is described 
by ideal fluid dynamics. For small departures from equilibrium, 
the system is described by dissipative fluid dynamics. 
The Navier-stokes equation is the fundamental equation in 
the non-relativistic  viscous hydrodynamics.   
However, in case of relativistic Navier-Stokes equation, 
due to the lack of an initial value formulation in first order 
hydrodynamics, signals can be transmitted with arbitrarily  
high speed thereby violating causality. The system in this case is 
described by parabolic equations. The first order theory 
has been extended by M\"uller \cite{Muller:67}; Israel and Stewart 
\cite{Israel:197679IS} by including the second order
gradient terms thereby preserving causality in the resulting relativistic 
hydrodynamics equations. 
The set of transport coefficients  are extended in the second order 
hydrodynamics and the resulting equations become hyperbolic.

The study of time dependent AdS/CFT correspondence in the context
of expansion of the RHIC, LHC fireball has been one of the 
important area of research. 
In the context of heavy-ion collisions, the basic features of the 
boost invariant evolution of the plasma have been discussed by 
studying the one dimensional expansion under the assumption that the 
system remains invariant under longitudinal 
boosts in the central rapidity region of the expanding plasma 
\cite{Bjorken:1983}. 
The gravity dual desription of the expansion
of the strongly coupled QGP has been very useful in understanding 
various aspects of both the boundary as well as bulk theory. 
The earlier work has also suggested that the gravity dual of 
RHIC fireball is a black hole \cite{Shuryak:2005SSZ, Nastase:2005}.  

In a pioneering work, Janik and Peschanski have used the AdS/CFT 
correspondence and Bjorken's boost invariance symmetry to study the 
dynamics of the strongly coupled QGP in 
non-viscous case \cite{Janik:2006JP1}.
Using conformal invariance of the gauge theory, energy-momentum 
conservation condition and hologhaphic renormalization method 
\cite{Haro:2001}, they have constructed the dual geometry from the given 
boundary data on the gauge theory side. Using Fefferman-Graham (FG) 
coordinates and solving the 
nonlinear Einstein's equations, the non-singular bulk geometry in 
large proper time has been obtained \cite{Janik:2006JP1}. The regularity 
of the bulk geometry has been an important aspect in determining the 
hydrodynamic quantities in the corresponding boundary theory. The 
boost invariant dual geometry has also been constructed  
by including conserved R-charge \cite{Bak:2006BJ}.
    
In view of the results from RHIC as well as LHC, it is important to 
include shear viscosity for the expanding plasma and analyse the 
corresponding dynamics. 
Including the shear viscosity in the context of relativistic first 
order hydrodynamics, the dual geometry has been constructed 
in Ref. \cite{Nakamura:2006NS}. 
Then the above study corresponding to Bjorken's one dimensional 
expansion has been extended by Sin, Nakamura and Kim to anisotropic 
three dimensional expansion of the plasma by considering Kasner 
space-time as local rest frame of the fluid \cite{Nakamura:2006SNK}. 
Though Kasner space-time is a curved space-time, the authors 
of Ref. \cite{Nakamura:2006SNK} have shown that under a well cotrolled 
approximation, it can be considered as the local rest frame of the 
anisotropically expanding fluid on Minkowski space-time. 
The corresponding gravity dual in large proper time regime has been 
obtained in first order gradient expansion using 
Fefferman-Graham coordinates \cite{Nakamura:2006SNK}. 
Subsequently,  
the form of the stress tensor within the framework 
of second order viscous hydrodynamics have been obtained and the 
associated transport coefficients have been explicitly determined in 
\cite{Baier:2007BRSSS, Sayantani:2007BHMR}. The relaxation time 
in second order viscous hydrodynamics has also been computed 
from the analysis of the regularity of the dual geometry 
\cite{Heller:2007HJ}.    

Though the Fefferman-Graham coordinates have been very  
useful for the gravity dual description, the regularity of the 
dual geometry in late time regime 
has been an important issue. This problem has been addressed
by working with Eddington-Finkelstein type coordinates. 
In the frame work of second order viscous hydrodynamics, it 
has been shown explicitly that the dual geometry is regular in the 
late time approximation and the regularity of the 
dual geometry determines the transport coefficients uniquely  
\cite{Heller:2008HSLSV, Kinoshita:2008KMNO, Kinoshita:2009PRL}.  
The dual geometry is regular except for the physical singularity 
at the origin which is covered by the event horizon of the bulk 
geometry. However, the concept of event horizon in a 
time dependent dual geometry is not entirely clear. Instead of an 
event horizon whose position is time dependent, it is more appropriate 
to consider a locally defined apparent horizon which is crucial in 
establishing that the dual geometry corresponds to a dynamical black hole
\cite{Kinoshita:2008KMNO, Mukund:2009}.

In this paper, we consider second order relativistic viscous hydrodynamics 
and study the anisotropic three dimensional expansion of the relativistic 
plasma in a time dependent background. The local rest frame (LRF) of the 
anisotropically expanding fluid is described by the time dependent 
Kasner space-time. We compute the dissipative part 
of the energy-momentum tensor in this set-up and obtain the explicit 
expressions for the components of the energy-momentum tensor in terms of 
Kasner parameters, pressure, energy density, shear viscosity, relaxation 
time and other transport coefficients in second order relativistic 
hydrodynamics. Then we solve the correspoding equations of state and 
hydrodynamic conservation equation for the conformal fluid and 
obtain the expressions for the time evolution of the energy density, 
pressure, temperature and entropy per unit comoving volume in terms of the 
transport coefficients in second order gradient expansion and the 
Kasner parameters. These expressions have not been obtained before. 
Subsequently, we have made an 
attempt to find the holographic dual of the anisotropically expanding 
plasma in the late time approximation in the context of time dependent 
AdS/CFT correspondence. We have 
tried to obtain the zeroth order solution for the corresponding 
dual five dimensional bulk space-time in 
Eddington-Finkelstein coordinates. We find that the  
zeroth order solution is an exact solution in the large proper time 
limit with constraints on the values of the Kasner parameters required 
for consistency.             

The paper is organized as follows: In section 2, we have considered 
Kasner space-time as the local rest frame 
of the fluid  and have studied the anisotropically expanding fluid in 
three dimensions thereby generalizing Bjorken's one dimensional 
expansion. Section 3 deals with the study of the evolution of the 
QGP in the context of second order viscous hydrodynamics and 
time dependent AdS/CFT correspondence. Using the conditions of 
conformal invariance and energy momentum conservation, we have obtained 
explicit expression for the energy density as a function of proper time 
in terms of Kasner parameters with  second order gradient expansion. 
The proper time dependence of the pressure, temperature and entropy per 
unit co-moving volume have also been obtained in terms of Kasner parameters.
Our results reduce to that of earlier discussions in the framework 
of one dimensional Bjorken expansion in appropriate limit of the Kasner 
parameters where the local rest frame is described by Minkowski space-time 
in terms of proper time and rapidity. In section 4, we have made a 
proposal for the gravity dual of the anisotropically expanding 
fluid using Eddington-Finkelstein type coordinates 
and have shown that the zeroth order solution is an exact solution of the 
five dimensional Einstein's equation in the large proper time limit 
with constraints on the Kasner parameters.  
We summarise and discuss the future perspective in section 5. 

%%%%%%%%%%%%%%%%%%%%%%%%%%%%%%%%%%%%%%%%%%%%%%%%%%%%%%%%%%%%%%%%%%%%
\section{Bjorken's hydrodynamics and Kasner spacetime}
\label{sec:Bjorken's hydrodynamics-Kasner spacetime}
\setcounter{equation}{0}
%%%%%%%%%%%%%%%%%%%%%%%%%%%%%%%%%%%%%%%%%%%%%%%%%%%%%

In this section, we consider the viscous hydrodynamics including second 
order gradient expansion terms and study the anisotropic three 
dimensional expansion of the fluid with Kasner spacetime as the 
local rest frame. For the one dimensional Bjorken expansion case,   
the local rest frame of the fluid is described by the proper time 
$\tau$ and rapidity $y$ and they are related to the cartesian 
coordinates in the following way,
\begin{eqnarray}
(X_{0},X_{1},X_{2},X_{3}) = 
(\tau \, \text{coshy},\tau \, \text{sinhy}, X_2, X_3)
\end{eqnarray}
Here the collision axis is taken along $X_{1}$ direction. 
The Minkowski metric in these coordinate is given by, 
\begin{eqnarray}
ds^{2} = - (d \tau)^{2} + \tau^{2}dy^{2}+(dX_{2})^2 + (dX_{3})^2 
\end{eqnarray}

We consider the generalization of Bjorken's one dimensional expansion 
to three dimensional expansion of the plasma in order to 
connect to the realistic description of the RHIC and LHC fireball
and for this, we consider Kasner space-time as the local rest
frame of the fluid \cite{Nakamura:2006SNK}. The metric is given by,
\begin{eqnarray}
ds^{2} = - (d \tau)^{2}+\tau^{2a}(dx_{1})^{2}+\tau^{2b}(dx_{2})^{2}+
\tau^{2c}(dx_{3})^{2}
\end{eqnarray}

Here $x_1, x_2, x_3$ are the comoving coordinates. $a, b, c$ are 
constants and are known as Kasner parameters. The Kasner parameters satisfy 
the conditions,
\begin{eqnarray}
a + b + c = 1, \hspace{20pt}a^2 + b^2 + c^2 = 1 
\end{eqnarray}

The above Kasner metric is an exact solution of vacuum Einstein's 
equation and it describes a homogeneous and anisotropic 
expansion of the Universe.  
The physical quantities are assumed to depend only on proper time 
$\tau$. The nonzero components of the affine connection 
for the Kasner metric are given by, 
\begin{eqnarray}
\begin{split}
\Gamma^{\tau}_{x_1 x_1} = a \tau^{2 a -1},\, \Gamma^{\tau}_{x_2 x_2}
= b \tau^{2 b -1},\, \Gamma^{\tau}_{x_3 x_3} = c \tau^{2 c -1},\\ 
\Gamma^{x_1}_{x_1 \tau} = \frac{a}{\tau}, \,
\Gamma^{x_2}_{x_2 \tau} = \frac{b}{\tau}, \,\Gamma^{x_3}_{x_3 \tau} = 
\frac{c}{\tau}
\end{split}
\end{eqnarray}

In Ref. \cite{Jaiswal:2017}, Kasner space-time has also been studied
to relate the anisotropic expansion with anisotropic hydrodynamics
(see Ref.\cite{Strickland:2014} for a review on anisotropic
hydrodynamics).

Since the gauge theory is conformal, from the tracelessness condition 
of the energy-momentum tensor, we obtain, 
\begin{eqnarray}
- T_{\tau \tau} + \frac{1}{\tau^{2 a}} \,T_{x_1 x_1} +  
\frac{1}{\tau^{2 b}} \,T_{x_2 x_2} +  \frac{1}{\tau^{2 c}}\,
T_{x_3 x_3} =0   
\end{eqnarray}
From the conservation of energy-momentum tensor \\
$\nabla_{\mu}\,T^{\mu\nu} = 0$ 
(where $\mu, \nu = \tau, x_1, x_2, x_3 $), one gets further relation 
among the components of $T_{\mu\nu}$ :
\begin{eqnarray}
\begin{split} 
\partial_{\tau}T_{\tau\tau} + \frac{(a+b+c)}{\tau} \,T_{\tau\tau} 
+ \frac{a}{\tau} \,\tau^{-2 a} \,T_{x_1 x_1}\\
 + \frac{b}{\tau} \,\tau^{-2 b} \,T_{x_2 x_2} +\frac{c}{\tau}\, 
\tau^{-2 c} \,T_{x_3 x_3} = 0
\end{split}
\end{eqnarray}

For $a=1, b=0$ and $c=0$, these equations reduce to that of the one 
dimensional expansion case with Minkowski space-time as the local rest 
frame of the fluid and the components of the energy-momentum tensor 
can be written in terms of a single function which is interpreted 
as the energy density as a function of time.     
 
In the relativistic viscous hydrodynamics, the energy-momentum tensor 
is given by, 
\begin{eqnarray}
T^{\mu\nu}=\epsilon\, u^{\mu}u^{\nu}+P\Delta^{\mu\nu}+\Pi^{\mu\nu}          
\end{eqnarray}
where $u^{\mu}$, $\epsilon$ and P are 4-velocity, 
the energy density and pressure respectively. $\Pi^{\mu\nu}$ 
represents the dissipative part. The dissipative part including 
second order gradient expansion terms is given by
\cite{Baier:2007BRSSS, Sayantani:2007BHMR} (we use the notations
of \cite{Baier:2007BRSSS}),
\begin{equation}
\begin{split}
\Pi^{\mu\nu}& = -\eta\sigma^{\mu\nu}+\eta\tau_{\pi}\left[ ^{\langle} 
D\sigma^{\mu\nu\rangle} +\dfrac{1}{3}\sigma^{\mu\nu}(\nabla\cdot u)
\right]\\
& \quad+ \kappa\left[R^{<{\mu\nu}>}-2u_{\alpha}R^{\alpha<{\mu\nu}>
\beta}u_{\beta}\right]\\
& \quad+\lambda_{1}\sigma^{<\mu}{{_\lambda}{\sigma}}^{{\nu>}
\lambda}+\lambda_{2}\sigma^{<\mu}{{_\lambda}{\Omega}}^{{\nu>}
\lambda}+\lambda_{3}\Omega^{<\mu}{{_\lambda}{\Omega}}^{{\nu>}
\lambda}
\end{split}
\end{equation}
where $\eta$ is the shear viscosity, $\tau_{\pi}$ is the relaxation time and 
$\kappa, \lambda_{1}, \lambda_{2}, \lambda_{3}$ are the other second order 
transport coefficients. For flat space, $\kappa$ term vanishes. In the 
first order hydrodynamics, only the first term in $\Pi^{\mu\nu}$ 
involving shear viscosity $\eta$ is relevant. The bulk viscosity is zero 
as we are considering a conformal fluid.   
\vskip 10pt 
The various terms appearing in $\Pi^{\mu\nu}$ are given by,  
\begin{equation*}
D\equiv u^{\mu}\nabla_{\mu}
\end{equation*}
\begin{equation*}
\Delta^{\mu\nu}=g^{\mu\nu}+u^{\mu}u^{\nu}
\end{equation*}
\begin{equation*}
{}^{\langle}\nabla^{\mu} u^{\nu\rangle} = 
\left( \Delta^{\mu
\lambda}\nabla_{\lambda}u^{\nu}+\Delta^{\nu\lambda}
\nabla_{\lambda}u^{\mu}-\dfrac{2}{3}\Delta^{\mu\nu}
\nabla_{\lambda}u^{\lambda}\right)
\end{equation*}
\begin{equation*}
\sigma^{\mu\nu} = 2 ^{\langle}\nabla^{\mu} u^{\nu\rangle}
\end{equation*}
\begin{equation*}
^{\langle}R^{\mu\nu\rangle} =
\dfrac{1}{2}\Delta^{\mu\alpha}
\Delta^{\nu\beta}(R_{\alpha\beta}+R_{\beta\alpha})-\dfrac{1}{3}
\Delta^{\mu\nu}\Delta^{\alpha\beta}R_{\alpha\beta}=R^{\left\langle{\mu\nu}
\right\rangle}
\end{equation*}
\begin{equation*}
R^{\alpha\langle\mu\nu\rangle\beta}=\dfrac{1}{2}\Delta^{\mu\sigma}
\Delta^{\nu\rho}(R^{\alpha\beta}_{\sigma\rho}+R^{\alpha\beta}_{\rho\sigma})-
\dfrac{1}{d-1}\Delta^{\mu\nu}\Delta^{\sigma\rho}R^{\alpha\beta}_
{\sigma\rho}
\end{equation*}
\begin{equation}
\Omega^{\mu\nu}=\dfrac{1}{2}\Delta^{\mu\alpha}\Delta^{\nu\beta}
(\nabla_{\alpha}u_{\beta}-\nabla_{\beta}u_{\alpha})
\end{equation}

where, $u^{\mu}=(1,0,0,0)$. $\Delta^{\mu\nu}$ is the projector on 
the spatial subspace, $\Omega$ is the vorticity term. In general, 
the bracket for a second rank tensor implies,  
\begin{equation}
^{\langle}A^{\mu\nu\rangle} \equiv  \dfrac{1}{2}\Delta^{\mu\alpha}
\Delta^{\nu\beta}(A_{\alpha\beta}+A_{\beta\alpha})-\dfrac{1}{d-1}
\Delta^{\mu\nu}\Delta^{\alpha\beta}A_{\alpha\beta}
\equiv A^{\left\langle{\mu\nu}\right\rangle}
\end{equation}  
%%%%%%%%%%%%%%%%%%%%%%%%%%%%%%%%%%%%%%%%%%%%%%%%%%%%%%%%%%%%%%%%
\section{Energy-momentum tensor in second order hydrodynamics}
\label{sec:Energy-momemntum tensor in second order hydrodynamics}
\setcounter{equation}{0}
%%%%%%%%%%%%%%%%%%%%%%%%%%%%%%%%%%%%%%%%%%%%%%%%%%%%%%%%%%%%%%%%%%%

In case of first order gradient expansion, the energy-momentum 
tensor in Kasner space-time is obtained as, 

\begin{equation}
T^{\mu \nu} = 
\begin{pmatrix} 
\epsilon (\tau) & 0 & 0 & 0 \\
0 & \dfrac{(P - \frac{2 \eta}{3 \tau} (3 a -1))}{\tau^{2 a}}  & 0 & 0 \\
0 & 0 & \dfrac{(P - \frac{2 \eta}{3 \tau} (3 b -1))}{\tau^{2 b}}  & 0  \\
0 & 0 & 0 & \dfrac{(P - \frac{2 \eta}{3 \tau} (3 c -1))}{\tau^{2 c}}  
\end{pmatrix}
\end{equation}

%\begin{eqnarray}
%T^{00} = \epsilon (\tau) \nonumber\\
%T^{11} =\tau^{-2 a} (P - \frac{2 \eta}{3 \tau} (3 a -1))\nonumber\\
%T^{22} =\tau^{-2 b} (P - \frac{2 \eta}{3 \tau} (3 b -1))\nonumber\\
%T^{33} = \tau^{-2 c} (P - \frac{2 \eta}{3 \tau} (3 c -1))
%\end{eqnarray} 

The corresponding equation of state and conservation condition are 
respectively given by, 
\begin{eqnarray} 
P & = \dfrac{\epsilon}{3} \\
\dfrac{d \epsilon}{d \tau} + \dfrac{4 \epsilon}{3 \tau} 
- \dfrac{4 \eta}{3 \tau^2} & = 0 
\end{eqnarray} 
As one can see, these equations are independent of the Kasner parameters.
\vskip 10pt

For the second order hydrodynamics, we obtain the explicit expressions 
for the components of the dissipative part ($\Pi^{\mu\nu}$) of the 
energy-momentum tensor in terms of Kasner parameters, which are given by,

\begin{eqnarray}
\Pi^{00} = 0 \nonumber\\
\Pi^{11}  = -\dfrac{2\eta(3a-1)}
{3\tau^{2a+1}}-\dfrac{4\eta \tau_{\pi}(3a-1)}{9\tau^{2a+2}}-
\dfrac{2\kappa(-1+a)a}{\tau^{2a+2}}
+\dfrac{4\lambda_1(9a^{2}-6a-1)}
{9\tau^{2a+2}} \nonumber \\
\Pi^{22}  = -\dfrac{2\eta(3b-1)}
{3\tau^{2b+1}}-\dfrac{4\eta \tau_{\pi}(3b-1)}{9\tau^{2b+2}}-
\dfrac{2\kappa(-1+b)b}{\tau^{2b+2}}
+\dfrac{4\lambda_1(9b^{2}-6b-1)}
{9\tau^{2b+2}} \nonumber \\
\Pi^{33}  = -\dfrac{2\eta(3c-1)}
{3\tau^{2c+1}}-\dfrac{4\eta \tau_{\pi}(3c-1)}{9\tau^{2c+2}}-
\dfrac{2\kappa(-1+c)c}{\tau^{2c+2}}
+\dfrac{4\lambda_1(9c^{2}-6c-1)}
{9\tau^{2c+2}}
\end{eqnarray}

One can check, for the one dimensional expansion of the fluid, corresponding 
to $a=1,b=0,c=0$, the above expressions become (written in 
a matrix form) \cite{Kinoshita:2008KMNO},
\begin{eqnarray}
\Pi^{\mu\nu}=-\eta
\begin{pmatrix}
    0  &   &   & \\
      &  \dfrac{4}{3}\tau^{-3}   & &\\
       &   &  -\dfrac{2}{3}\tau^{-1}  & \\
       & & & -\dfrac{2}{3}\tau^{-1}
\end{pmatrix}
+ (\eta \tau_{\pi}-\lambda_1)
\begin{pmatrix}
     0  &   &   &      \\
   &  -\dfrac{8}{9}\tau^{-4}   & &\\
       &   &  \dfrac{4}{9}\tau^{-2}  & \\
       & & & \dfrac{4}{9}\tau^{-2}
\end{pmatrix} 
\end{eqnarray}
%\begin{eqnarray}
%\Pi^{00} = 0 \nonumber\\
%\Pi^{11}  = -\dfrac{4\eta\tau^{-3}}
%{3}-(\eta \tau_{\pi}-\lambda_1)\dfrac{8}{9}\tau^{-4}\nonumber \\
%\Pi^{22}  = \dfrac{2\eta\tau^{-3}}
%{3}+(\eta \tau_{\pi}-\lambda_1)\dfrac{4}{9}\tau^{-4}\nonumber \\
%\Pi^{33}  = \dfrac{2\eta\tau^{-3}}
%{3}+(\eta \tau_{\pi}-\lambda_1)\dfrac{4}{9}\tau^{-4}\nonumber \\
%\end{eqnarray}

Next, for the three dimensional expansion of the fluid, we get the 
components of the energy momentum tensor as:
\begin{equation*}
T^{00}= \epsilon(\tau)
\end{equation*}
\begin{equation*}
T^{11}= \dfrac{P(\tau)}{\tau^{2a}}-\dfrac{2\eta(3a-1)}{3\tau^{2a+1}}-
\dfrac{4\eta \tau_{\pi}(3a-1)}{9\tau^{2a+2}}+
\dfrac{2\kappa(1-a)a}{\tau^{2a+2}} 
+\dfrac{4\lambda_1(9a^{2}-6a-1)}
{9\tau^{2a+2}}                
\end{equation*}
\begin{equation*}
T^{22} = \dfrac{P(\tau)}{\tau^{2b}}-\dfrac{2\eta(3b-1)}{3\tau^{2b+1}}-
\dfrac{4\eta \tau_{\pi}(3b-1)}{9\tau^{2b+2}}+
\dfrac{2\kappa(1-b)b}{\tau^{2b+2}} 
+\dfrac{4\lambda_1(9b^{2}-6b-1)}{9\tau^{2b+2}}
\end{equation*}
\begin{align}
T^{33} = \dfrac{P(\tau)}{\tau^{2c}}-\dfrac{2\eta(3c-1)}{3\tau^{2c+1}}-
\dfrac{4\eta \tau_{\pi}(3c-1)}{9\tau^{2c+2}}+\dfrac{2\kappa(1-c)c}{\tau^{2c+2}} 
+\dfrac{4\lambda_1(9c^{2}-6c-1)}
{9\tau^{2c+2}}
\end{align}               

We have obtained these expressions after putting the Kasner conditions, 
\begin{equation}
\sum_{i}a_{i}=1,\hspace{20pt}\sum_{i}a_{i}^{2}=1 \qquad (a_{i}= a, b, c)
\end{equation}
  
The hydrodynamic conservation equation $\nabla_{\mu}T^{\mu\nu}=0$ becomes,
\begin{align}
\dfrac{d\epsilon}{d\tau} &=-\dfrac{(a+b+c)\epsilon}{\tau}-\dfrac{(a+b+c)P}
{\tau}+\dfrac{2\eta}{\tau^{2}}\left( (a^{2}+b^{2}+c^{2})-\dfrac{1}{3}
(a+b+c)^{2}\right)\nonumber\\
& \qquad  -\dfrac{\eta\tau_{\pi}}{\tau^{3}}\left\lbrace-2(a^{2}+b^{2}+c^{2})+
\dfrac{2}{3}(a^{2}+b^{2}+c^{2})(a+b+c)+\dfrac{2}{3}(a+b+c)^{2}-
\dfrac{2}{9}(a+b+c)^{3}\right\rbrace \nonumber\\
& \qquad -\dfrac{\lambda_{1}}{3\tau^{3}}\left\{(a-\dfrac{1}{3}(a+b+c))^{2}
(8a-4b-4c)+(b-\dfrac{1}{3}(a+b+c))^{2}(8b-4a-4c)\right.\nonumber\\
& \qquad \left.{}+(c-\dfrac{1}{3}(a+b+c))^{2}(8c-4a-4b)\right\}\nonumber\\ 
& \qquad - \dfrac{\kappa}{\tau^{3}}\left\{(a+b+c)(a^{2}+b^{2}+c^{2})-
(a^{2}+b^{2}+c^{2})-\dfrac{1}{3}(a+b+c)^{3}+\dfrac{1}{3}(a+b+c)
\right.\nonumber\\
& \qquad \left.{}-2\left((a^{3}+b^{3}+c^{3})-2(a^{2}+b^{2}+c^{2})+(a+b+c)
\right)\right\} 
\end{align}

%\begin{align}
%\dfrac{d\epsilon}{d\tau} &=-\dfrac{(a+b+c)\epsilon}{\tau}-\dfrac{(a+b+c)P}
%{\tau}\nonumber\\
%& \qquad+\dfrac{2\eta}{\tau^{2}}\left( (a^{2}+b^{2}+c^{2})-\dfrac{1}{3}
%(a+b+c)^{2}\right)\nonumber\\
%& \qquad  -\dfrac{\eta\tau_{\pi}}{\tau^{3}}\left\lbrace-2(a^{2}+b^{2}+c^{2})
%+\dfrac{2}{3}(a+b+c)^{2}\right.\nonumber\\
%& \qquad \left.{}+\dfrac{2}{3}(a^{2}+b^{2}+c^{2})(a+b+c)-
%\dfrac{2}{9}(a+b+c)^{3}\right\rbrace \nonumber\\
%& \qquad -\dfrac{\lambda_{1}}{3\tau^{3}}\left\{(a-\dfrac{1}{3}(a+b+c))^{2}
%(8a-4b-4c)\right.\nonumber\\
%& \qquad \left.{}+(b-\dfrac{1}{3}(a+b+c))^{2}(8b-4a-4c)\right.\nonumber\\
%& \qquad \left.{}+(c-\dfrac{1}{3}(a+b+c))^{2}(8c-4a-4b)\right\}\nonumber\\ 
%& \qquad - \dfrac{\kappa}{\tau^{3}}\left\{(a+b+c)(a^{2}+b^{2}+c^{2})\right.
%\nonumber\\
%& \qquad \left.{}-(a^{2}+b^{2}+c^{2})-\dfrac{1}{3}(a+b+c)^{3}+
%\dfrac{1}{3}(a+b+c)\right.\nonumber\\
%& \qquad \left.{}-2\left((a^{3}+b^{3}+c^{3})-2(a^{2}+b^{2}+c^{2})
%\right)\right.\nonumber\\
%& \qquad \left.{}-2(a+b+c)\right\} 
%\end{align}
The equation of state and the conservation law become 
(using Kasner conditions),

\begin{align}
P = \dfrac{\epsilon}{3}
\end{align}
\begin{eqnarray}
\dfrac{d\epsilon}{d\tau}+\dfrac{4\epsilon}{3\tau} =\dfrac{4\eta}
{3\tau^{2}}+\dfrac{8\eta\tau_{\pi}}{9\tau^{3}} + 
\dfrac{2\kappa(-1+a^3+b^3+c^3)}{\tau^{3}}
-\dfrac{\lambda_{1}(-7+9(a^{3}+b^{3}+c^{3}))} 
{9\tau^{3}}
\end{eqnarray}

As one can see, the equation of state is independent of the Kasner 
parameters, but the energy-momentum conservation law does depend 
on the Kasner parameters.
  
From the conformal invariance of the fluid, the proper-time dependence 
of the transport coefficients are given by 
\begin{eqnarray}
\eta=\epsilon_{0}\,\eta_{0} \left( \dfrac{\epsilon}{\epsilon_{0}}\right)^{3/4},
\hspace{10pt}\tau_{\pi}=\tau_{\pi}^{0} \left( \dfrac{\epsilon}
{\epsilon_{0}}\right)^{-1/4}\hspace{10pt},\nonumber\\
\lambda_{1}=\epsilon_{0}\,
\lambda_{1}^{0} \left( \dfrac{\epsilon}{\epsilon_{0}}\right) ^{1/2},
\hspace{10pt} 
\kappa=\epsilon_{0}\,\kappa_{0} 
\left( \dfrac{\epsilon}{\epsilon_{0}}\right)^{1/2}
\end{eqnarray}
where  $\epsilon_{0}, \eta_{0}, \tau_{\pi}^{0}, \lambda_{1}^{0}, \kappa_{0}$ 
are constants.

The solution of the equation for energy density $\epsilon (\tau)$ 
is obtained as, 

\begin{align}
\dfrac{\epsilon(\tau)}{\epsilon_{0}}=\tau^{-4/3}-2\dfrac{\eta_{0}}{\tau^{2}}+
\left [ \frac{3\eta_{0}^{2}}{2}
+ \frac{\lambda_{1}^{0}}{3} (-7+9 a^{3}+9 b^{3}+9 c^{3})\right.\nonumber\\
\left.{}- \frac{2 \eta_{0}\tau_{\pi}^{0}}{3} - \frac{3 \kappa_{0}}{2}  
(-1+a^{3}+b^{3}+c^{3})
\right ] \tau^{-8/3}+...
\end{align}

Denoting the term in the above square bracket as 
\begin{eqnarray}
\left [ \frac{3\eta_{0}^{2}}{2}
+ \frac{\lambda_{1}^{0}}{3} (-7+9 a^{3}+9 b^{3}+9 c^{3})- 
\frac{2 \eta_{0}\tau_{\pi}^{0}}{3}\right.\nonumber\\
\left.{} - \frac{3 \kappa_{0}}{2}  
(-1+a^{3}+b^{3}+c^{3})
\right ] = \tilde\epsilon_{0}^{(2)},
\end{eqnarray}

the solution for $\epsilon (\tau)$ can be written as, 
\begin{equation}
\dfrac{\epsilon(\tau)}{\epsilon_{0}}=\tau^{-4/3}-2\eta_{0}\tau^{-2}+
\tilde\epsilon_{0}^{(2)}\tau^{-8/3}+...,
\end{equation}
%where $\tilde\epsilon_{0}^{(2)}$ is given as, 
%\begin{equation}
%\tilde\epsilon_{0}^{(2)}=\dfrac{9\eta_{0}^{2} - 
%9\kappa_{0}(-1+a^{3}+b^{3}+c^{3})
%+2\lambda_{1}^{0}(-7+9 a^{3}+9 b^{3}+9 c^{3})-4\eta_{0}\tau_{\pi}^{0}}{6}
%\end{equation}
For $a=1,b=0,c=0$, the above solution reduces to that of the 
one dimensional expansion case (where $\kappa = 0$ in flat space)
in second order hydrodynamics \cite{Baier:2007BRSSS}.
\vskip 10pt 
Now we write the components of 
the energy momentum tensor in terms of energy density and the 
expressions are given by,
\begin{equation*}
T_{00}/\epsilon_{0}=\tau^{-4/3}-2\eta_{0}\tau^{-2}+\tilde\epsilon_{0}^{(2)}
\tau^{-8/3}+...
\end{equation*}
\begin{equation*}
\begin{split}
\dfrac{T_{11}}{\epsilon_{0}\tau^{2a}} & =\dfrac{1}{3}\tau^{-4/3}-
2\eta_{0}a\tau^{-2}+
\eta_{0}^{2}(3a-\dfrac{1}{2})\tau^{-8/3}-
\dfrac{\kappa_{0}}{2}((a^{3}+b^{3}+c^{3})+(4a^{2}-4a-1))\tau^{-8/3}\\
& \quad -\dfrac{2}{9}\eta_{0}\tau_{\pi}^{0}(6a-1)
\tau^{-8/3}+\dfrac{\lambda_{1}^{0}}{9}
(9(a^{3}+b^{3}+c^{3})+(36a^{2}-24a-11))\tau^{-8/3} +...
\end{split}
\end{equation*}
\begin{equation*}
\begin{split}
\dfrac{T_{22}}{\epsilon_{0}\tau^{2b}} & =\dfrac{1}{3}\tau^{-4/3}-
2\eta_{0}b\tau^{-2}+
\eta_{0}^{2}(3b-\dfrac{1}{2})\tau^{-8/3}-
\dfrac{\kappa_{0}}{2}((a^{3}+b^{3}+c^{3})+(4b^{2}-4b-1))\tau^{-8/3}\\
& \quad -\dfrac{2}{9}\eta_{0}\tau_{\pi}^{0}(6b-1)
\tau^{-8/3}+\dfrac{\lambda_{1}^{0}}{9}
(9(a^{3}+b^{3}+c^{3})+(36b^{2}-24b-11))\tau^{-8/3}+...
\end{split}
\end{equation*}
\begin{equation}
\begin{split}
\dfrac{T_{33}}{\epsilon_{0}\tau^{2c}} & =\dfrac{1}{3}\tau^{-4/3}-
2\eta_{0}c\tau^{-2}+
\eta_{0}^{2}(3c-\dfrac{1}{2})\tau^{-8/3}-\dfrac{\kappa_{0}}{2}((a^{3}+b^{3}+c^{3})+(4c^{2}-4c-1))\tau^{-8/3}\\
& \quad -\dfrac{2}{9}\eta_{0}\tau_{\pi}^{0}(6c-1)
\tau^{-8/3}+\dfrac{\lambda_{1}^{0}}{9}
(9(a^{3}+b^{3}+c^{3})+(36c^{2}-24c-11))\tau^{-8/3}+...
\end{split}
\end{equation}

%If $a=1,b=0,c=0$, then the above equation reduces to the one dimensional 
%case.\\
From Stefan-Boltzmann's law, where $\epsilon \propto T^{4}$,
we obtain the proper time dependence of the temperature T as,  
\begin{eqnarray}
T (\tau) = \epsilon_{0}^{1/4}\left( \dfrac{1}{\tau^{1/3}}-
\dfrac{\eta_{0}}{2\tau}
+\dfrac{3\kappa_{0}(-1+a^{3}+b^{3}+c^{3})}{8\tau^{5/3}}\right.\nonumber\\
\left.{}+\dfrac{\lambda_{1}^{0}
(-7+9 a^{3}+9 b^{3}+9 c^{3})}{12\tau^{5/3}}
-\dfrac{\eta_{0}\tau_{\pi}^{0}}{6\tau^{5/3}}+...\right) 
\end{eqnarray}
It is useful to reexpress the conservation law of energy-momentum tensor as\\
\begin{eqnarray}
\dfrac{d(\sqrt{g}\epsilon)}{d\tau}+\dfrac{d\sqrt{g}}{d\tau}P =\dfrac{4}{3} 
\dfrac{\sqrt{g}\eta}{\tau^{2}}+\dfrac{8}{9} \dfrac{\sqrt{g}\eta\tau_{\pi}}
{\tau^{3}}
+\dfrac{2\sqrt{g}\kappa(-1+a^3+b^3+c^3)}{\tau^{3}}\nonumber\\
-\dfrac{4\sqrt{g}\lambda_{1}(-7+9a^3+9b^3+9c^3)}{9\tau^{3}}
\end{eqnarray}
where $\sqrt{g}=\tau$ is the volume element in the co-moving coordinate. 
Using the thermodynamic relation  $dE+PdV=TdS$, the above equation can be 
expressed as\\
\begin{eqnarray}
T\dfrac{d(\sqrt{g} s)}{d\tau} =\dfrac{4}{3} \dfrac{\sqrt{g}\eta}{\tau^{2}}+\dfrac{8}{9} \dfrac{\sqrt{g}\eta\tau_{\pi}}{\tau^{3}}
+\dfrac{2\sqrt{g}
\kappa(-1+a^3+b^3+c^3)}{\tau^{3}}\nonumber\\
 -\dfrac{4\sqrt{g}\lambda_{1}(-7+9a^3+9b^3+9c^3)}{9\tau^{3}}
\end{eqnarray}
where $s$ is the entropy density and $\tau s=\sqrt{g}s=S $ is the entropy 
per unit co-moving volume. Integrating the above equation and using the 
proper time dependence of the temperature, the entropy per unit 
co-moving volume (as a function of proper time $\tau$) is obtained as,
%\\
\begin{eqnarray}
S(\tau)=\dfrac{4}{3}\int_{0}^{\tau}d\tau\dfrac{\sqrt{g}\eta}{\tau^{2}T}
+\dfrac{8}{9}\int_{0}^{\tau}d\tau\dfrac{\sqrt{g}\eta\tau_{\pi}}{\tau^{3}T}
\nonumber \\
+2\int_{0}^{\tau}d\tau\dfrac{\sqrt{g}\kappa(-1+a^3+b^3+c^3)}
{\tau^{3}T} \nonumber \\
-\dfrac{4}{9}\int_{0}^{\tau}d\tau\dfrac{\sqrt{g}\lambda_{1}(-7+9a^3
+9b^3+9c^3)}{9\tau^{3}T}\\
=\epsilon_{0}^{3/4}\left\{1-\dfrac{3\eta_{0}}{2}\tau^{-2/3}+
\dfrac{3\eta_{0}^{2}}{4}\tau^{-4/3}-
\dfrac{\eta_{0}\tau_{\pi}^{0}}{2}\tau^{-4/3}\right.\nonumber\\
\left.{} 
+\dfrac{\lambda_{1}^{0}(-7+9a^3+9b^3+9c^3)}{4}\tau^{-4/3}\right.\nonumber\\
\left.{} -\dfrac{9\kappa_{0}(-1+a^3+b^3+c^3)}{8}
\tau^{-4/3}+0(\tau^{-2}) \right\}
\end{eqnarray}
%We have checked that for $a=1, b=0, c=0$, the above expressions reduce to 
%the one dimensional case of 
%Ref. \cite{Kinoshita:2008KMNO, Kinoshita:2009PRL}.
%\vskip 1pt
As one can see, our expressions for energy density, temperature and 
entropy as a function of proper time $\tau$ in second order hydrodynamics 
in Kasner space time depend on the Kasner parameters $a, b$ and $c$.

%%%%%%%%%%%%%%%%%%%%%%%%%%%%%%%%%%%%%%%%%%%%%%%%%%%%%%%%%%%%%%%%%%%%
\section{A proposal for the gravity dual of anisotropic expansion}
\label{sec:Gravity dual description of anisotropic expansion}
\setcounter{equation}{0}
%%%%%%%%%%%%%%%%%%%%%%%%%%%%%%%%%%%%%%%%%%%%%%%%%%%%%%%%%%%%%%%%%%%

In this section, we shall discuss the holographic dual of the 
anisotropically expanding fluid in the late time approximation using 
Eddington-Finkelstein (EF) coordinates. 
For the one dimensional expansion case, the gravity dual geometry has been 
obtained in \cite{Nakamura:2006SNK} by using the 
Fefferman-Graham (FG) Coordinates. The five dimensional asymptotically AdS 
metric in FG coordinates is given by, 
\begin{equation}
ds^{2}=\dfrac{g_{\mu\nu}dx^{\mu}dx^{\nu}+dz^{2}}{z^{2}}
\end{equation}
where $x^{\mu}=(\tau, x_1, x_2, x_3)$.  
$g_{\mu\nu}$ is the four-dimensional metric which depends on 
both $\tau$ and $z$ and it is expanded with respect to $z$ 
as \cite{Haro:2001},
\begin{equation}
g_{\mu\nu}(\tau,z)=g_{\mu\nu}^{(0)}(\tau)+z^{2}g_{\mu\nu}^{(2)}
(\tau)+z^{4}g_{\mu\nu}^{(4)}(\tau)+z^{6}g_{\mu\nu}^{(6)}(\tau)+....
\end{equation}
Here, $g_{\mu\nu}^{(0)}$ is the gauge theory metric  
on the boundary. In our case, $g_{\mu\nu}^{(0)}$ corresponds 
to the Kasner metric.  
$g_{\mu\nu}^{(2)} = 0$  as the four dimensional 
metric is Ricci flat. 
$g_{\mu\nu}^{(4)}$ is proportional to the boundary energy-momentum tensor, 
namely,
\begin{equation}
g_{\mu\nu}^{(4)} = {\it const} \left\langle T_{\mu\nu}\right\rangle 
\end{equation} 
One obtains the higher order terms in the expansion of 
$g_{\mu\nu}(\tau, z)$ by solving the five dimensional bulk Einstein's 
equation with a negative cosmological constant recursively:  
\begin{equation}
R_{MN}-\dfrac{1}{2}G_{MN}R-6 G_{MN}=0
\end{equation}
where $G_{MN}$, $R_{MN}$ and $R$ correspond to the metric, Ricci tensor 
and Ricci scalar respectively in the five dimensional theory.
After solving the Einstein’s equation recursively, the late time 
5D bulk geometry can be obtained in a compact form.
The dual metric has been obtained by Sin, Nakamura and Kim 
with first order gradient expansion terms and is given by
\cite{Nakamura:2006SNK}, 
\begin{eqnarray}
ds^{2}=\dfrac{1}{z^{2}} \left\lbrace  -\dfrac{{\left( 1-\dfrac{\epsilon z^{4}}
{3}\right) }^{2}}{1+\dfrac{\epsilon z^4}{3}} d\tau^{2} 
+\left( 1+\dfrac{\epsilon z^4}{3}\right)\right.\nonumber\\
\left.{}\sum_{i=1}^{3} 
{\left( \dfrac{1+\dfrac{\epsilon z^{4}}{3}}{1-\dfrac{\epsilon z^4}{3}} 
\right)}^{(1-3a_{i})\gamma} \tau ^{2a_{i}} (dx^{i})^{2}  \right\rbrace
+\dfrac{dz^{2}}{z^{2}}
\end{eqnarray}
where,
\begin{eqnarray}
\epsilon(\tau)=\dfrac{\epsilon_{0}}{\tau^{4/3}}-\dfrac{2\eta_{0}}{\tau^{2}}, 
\hspace{15pt}
\gamma=\dfrac{\eta_{0}}{\epsilon_{0}\tau^{2/3}},\nonumber\\
 a_i (i = 1,2,3) \equiv a, b, c.
\end{eqnarray}

The above 5D metric in FG coordinate is a solution in the late time 
regime and is correct only upto order 
$\gamma$ ($\gamma\propto {\tau}^{-2/3}$) \cite{Nakamura:2006SNK}.
\vskip 5pt
  
Though the holographic dual of Bjorken flow has been described very well 
by using the Fefferman-Graham (FG) coordinate, it is difficult to define the 
location of the event horizon in a time dependent geometry in this 
coordinate. In a very interesting paper, the authors of 
Ref. \cite{Kinoshita:2008KMNO} (see also \cite{Kinoshita:2009PRL}) 
have proposed the dual geometry in late time expansion by using 
Eddington-Finkelstein (EF) coordinates in the one dimensional 
expansion case and have also computed the location
of the apparent horizon (boundary between the trapped and untrapped 
region). The regularity of the dual geometry  
has been shown to all orders and that in turn, determines the transport 
coefficients uniquely \cite{Kinoshita:2008KMNO, Kinoshita:2009PRL}.   
Eddington Finkelstein coordinates have also been used before to construct 
the dual geometry which is regular except at the origin 
\cite{Sayantani:2007BHMR, Heller:2008HSLSV}. 

In analogy with the one dimensional expansion of the fluid 
\cite{Kinoshita:2008KMNO, Kinoshita:2009PRL}, we propose the following 
parametrization for the dual geometry in the late time regime 
corresponding to the three dimensional expansion case with Kasner 
space-time as the local rest rame of the fluid:
\begin{align}
%\begin{eqnarray}
  d s^2 = - r^2 P d \tau^2 + 2 d \tau d r +  
  r^2 \tau^{2 a} e^{2 Q 
-2 R} \left ( 1 + \frac{1}{u \tau^{2/3}}\right )^2 d x_1^2 \nonumber\\
+ \ r^2 \tau^{2 b} e^R d x_2^2 \  + \ r^2 \tau^{2 c} e^R d x_3^2
%\end{eqnarray}
\end{align}  
where $r$ is the fifth dimension. The variable $u$ is defined as 
$u = r \tau^{1/3}$ and the 
the late time approximation is taken in an expansion  
in $\tau^{-2/3}$  keeping $u$ fixed. $a, b, c$ are Kasner parameters 
and $P, Q, R$ are functions of $u$ and $\tau$. 
The boundary conditions correspond to $P \rightarrow 1$, $Q \rightarrow
0$ and $R \rightarrow 0$ as $r \rightarrow \infty$ (corresponds to the 
spatial boundary with $r$ as the fifth dimension). 
\vskip 5pt
With these boundary conditions, 
the 5D bulk metric in the limit as $r \rightarrow \infty$ becomes, 

\begin{eqnarray} 
d s^2 |_{r \rightarrow \infty} = r^2 \left [- (d \tau)^{2} + 
\tau^{2a}(dx_{1})^{2}+\tau^{2b}(dx_{2})^{2}\right.\nonumber\\
\left.{} + \tau^{2c}(dx_{3})^{2} 
\right ] + 2 d \tau d r
\end{eqnarray}
where the quantity inside the square bracket is the boundary 
four dimensional Kasner metric on the local rest frame of the fluid.
This is in analogy with the one dimensional expansion with the Minkowski
metric on the LRF of the fluid \cite{Kinoshita:2009PRL}.
The 4D part of the proposed dual bulk metric have been taken to be 
diagonal since the 4D boundary energy-momentum tensor is diagonal.     
The parameters $P, Q, R$ are expanded in powers of $\tau^{-2/3}$ as 
\cite{Kinoshita:2008KMNO, Kinoshita:2009PRL}, 
\begin{eqnarray} 
P(\tau, u) & = P_0 (u) + P_1 (u) \tau^{-2/3} + P_2 (u) \tau^{-4/3} 
+ ...  \nonumber \\
Q(\tau, u) & = Q_0 (u) + Q_1 (u) \tau^{-2/3} + Q_2 (u) \tau^{-4/3} 
+ ...  \nonumber \\
R(\tau, u) & = R_0 (u) + R_1 (u) \tau^{-2/3} + R_2 (u) \tau^{-4/3} 
+ ... 
\end{eqnarray}
where $P_n$, $Q_n$ and $R_n$ are obtained by solving the 5D Einstein's 
equation order by order in late time regime with the boundary conditions
as mentioned above.
The zeroth order solution is given by (we have set the integration 
constant to zero),
\begin{equation}
P_0(u) = 1 - \dfrac{w^4}{u^4}, \qquad Q_0 = 0, \qquad R_0 = 0 
\end{equation} 
where, $w$ is a constant. We would like to clarify that the 
corresponding zeroth order metric given by,  
\begin{eqnarray}
d s^2 = - r^2 \left ( 1 - \dfrac{w^4}{u^4}\right )  d \tau^2 + 
2 d \tau d r 
+ 
r^2 \tau^{2 a} \left ( 1 + \dfrac{1}{u \tau^{2/3}}\right )^2 d x_1^2 
\nonumber\\ + r^2 \tau^{2 b} d x_2^2 + r^2 \tau^{2 c} d x_3^2
\end{eqnarray}
is an exact solution of the 5D Einstein's equation in the large $\tau$ 
limit. Consistency of the above solution puts a constraint on the values 
of the Kasner parameters, namely $a=1$, $b=0$ and $c=0$ so that 
the zeroth order metric becomes an exact solution in the large $\tau$ limit. 
There can be other choices for $a, b$ and $c$, but they do not 
satisfy the Kasner conditions, $a + b + c =1$ and $a^2 + b^2 + c^2 = 1$. 
%Since the 5D metric gets corrected in a $\tau^{-2/3}$ expansion in 
%subsequent orders, there 
%is no need for the zeroth order solution to be an exact solution 
%for all $\tau$.  
At zeroth order, the term $\dfrac{1}{u \tau^{2/3}}$ in the bracket 
in the coefficient of $d x_1^2$ is ignored \cite{Kinoshita:2008KMNO}. 
This term was introduced so that the zeroth order metric in the one 
dimensional expansion case reduces to an exact AdS metric (through 
a coordinate transformation) in the $w \rightarrow 0$ limit 
\cite{Kinoshita:2008KMNO, Kinoshita:2009PRL}. 

The corresponding Kretschmann scalar for our zeroth order metric 
is obtained as, 

\begin{eqnarray} 
R_{MNKL} \, R^{MNKL} = 
\left [ 40 + \dfrac{72 w^8}{u^8} + \dfrac{32 (a + b + c)}{u \tau^{2/3}}
\right. \nonumber\\
\left.{}+ \dfrac{8 (a b + b c + c a)}{u^2 \tau^{4/3}} + 
\dfrac{16 (a^2 + b^2 + c^2)}{u^2 \tau^{4/3}} \right ] 
\end{eqnarray}
where, $M, N, K, L$ are the indices corresponding to the 5D metric. 
We have checked that the above expression reduces exactly 
to the zeroth order computation of the Kretschmann scalar  
in the one dimensional expansion case \cite{Kinoshita:2008KMNO} in 
appropriate limit of the Kasner parameters. 
After putting the Kasner conditions ($a + b + c = 1$, 
$a^2 + b^2 + c^2 = 1$ and $ab + bc + ca =0$ by virtue of the Kasner
conditions), the above expression for the Kretschmann scalar becomes 
independent of the Kasner parameters and is given by,   
\begin{equation} 
R_{KLMN} \, R^{KLMN} = 
8 \left ( 5 + \dfrac{9 w^8}{u^8} \right ) + O(\tau^{-2/3}), 
\end{equation}
where the physical singularity is at $u = 0$. This matches with the 
result of ref. \cite{Kinoshita:2008KMNO} in zeroth order. 
The Ricci scalar of the zeroth order metric is given by, 
\begin{equation}
R = - 20 - \dfrac{8 (a + b + c)}{u \tau^{2/3}}
 = - 20 - \dfrac{8}{u \tau^{2/3}}
\end{equation} 

A detailed analysis regarding the location of the apparent horizon 
of the dual geometry in an expansion in $\tau^{-2/3}$ 
has been done in Ref. \cite{Kinoshita:2008KMNO}, where the position 
of the apparent horizon has been expanded as, 
\begin{equation} 
u_H = u_0 + u_1 \tau^{-2/3} + u_2 \tau^{-4/3} + .... 
\end{equation} 
The zeroth order result gives $u_0 = w$. The higher order results 
have also been obtained by using the regularity condition of the 
dual geometry \cite{Kinoshita:2008KMNO, Kinoshita:2009PRL}, 
showing that there is indeed 
an apparent horizon. The existence of the apparent horizon also 
means that there is an event horizon so that the physical 
singularity at the origin is covered and it is not a naked singularity. 
(see Ref. \cite{Mukund:2009} for a rigorous discussion on 
the location of the apparent horizon and event horizon). 
We have not done a detailed analysis for the higher orders in 
expansion in $\tau^{-2/3}$ in the three 
dimensional expansion case. The volume element of the apparent horizon 
of the dual geometry in our three dimensional case (after putting 
Kasner condition) is given by, 
\begin{equation} 
vol. = r^3 \tau e^Q \left ( 1 + \dfrac{1}{r \tau} \right )
\end{equation}
which can be computed at $u = u_H$ in an expansion in $\tau^{-2/3}$.
We have checked that our expression for the zeroth order and first 
order terms in entropy density computed from 
hydrodynamics in the limit of $a=1, b=0, c=0$ match with the 
results computed from the volume element of the apparent horizon of the 
dual geometry in late time regime with appropriate normalization factor 
\cite{Kinoshita:2008KMNO}. It has been noticed that at second order, 
there is a mismatch between the geometrical and hydrodynamical 
entropy density in the one dimensional expansion case.
This does not imply any physical inconsistency 
and there can be several intricate reasons for this 
\cite{Kinoshita:2008KMNO}.      
%%%%%%%%%%%%%%%%%%%%%%%%%%%%%%%%%%%%%%%%%%%%%%%%%%%%%%%%%%%%%%%%%%%%%
\section{Summary and discussion}
\label{sec:Summary and discussion}
\setcounter{equation}{0}
%%%%%%%%%%%%%%%%%%%%%%%%%%%%%%%%%%%%%%%%%%%%%%%%%%%%%%%%%%%%%%%%%%
%%%%%%%%%%%%%%%%%%%%%%%%%%%%%%%%%%%%%%%%%%%%%%%%%%%%%%%%%%%%%%%%%%%%

In this paper, we have studied the three dimensional anisotropic 
expansion of a conformal fluid by using Kasner space-time as the 
local rest frame of the fluid as an example of time dependent 
AdS/CFT correspondence. 
We have considered relativistic viscous hydrodynamics to second 
order in gradient expansion and have obtained the expressions 
for energy density, temperature 
and the components of the energy momentum tensor in terms of 
Kasner parameters and the transport coefficients in the late time regime. 
We have also obtained 
the entropy density per unity rapidity from the hydrodynamics 
side. In analogy with 
one dimensional expansion case \cite{Kinoshita:2008KMNO, 
Kinoshita:2009PRL}, we have made a proposal 
for the 5-dimensional dual geometry in the large proper time approximation
using Eddington-Finkelstein coordinates in the three dimensional 
expansion case with the boundary metric 
as the 4D Kasner space-time. 
We find that the zeroth order metric is an exact solution in the large 
proper time limit with contraints on the Kasner parameters. 
The zeroth order computation agrees with the one dimensional case. 
The corresponding Kretshmann scalar has been computed and is found 
to be regular except for the physical singularity at the origin $u = 0$.  
We plan to make a detailed analysis of the regularity of the 
dual geometry and as well as the location of the apparent horizon 
at higher orders in an expansion in $\tau^{-2/3}$ in future.
 
Though the focus has been on the applications of fluid dynamics 
near local equilibrium of the system by using the gradient expansion, 
it is important to explore whether the applicability can also be 
extended to fluid dynamics far from local equilibrium 
\cite{Romatschke:2017FLE}.
This issue has opened up a new direction called 
"resurgence" giving rise to hydrodynamic attractor solutions. 
Resurgence theory 
suggests that the gradient series becomes divergent but is Borel 
summable giving rise to hydrodynamic attractor 
solutions \cite{Heller:2015}. It is also applicable to large gradients.
It will be interesting to study the 
attractor solution in the present case of three dimensional anisotropic
expansion of the fluid with second and higher order viscous hydrodynamics 
(see Ref.\cite{Jaiswal:2019} for related discussion). 
To conclude, the present study of QGP dynamics with second order 
relativistic viscous hydrodynamics and anisotropic expansion of the fluid 
using Kasner space-time is expected to provide a better understanding 
of the physics of early universe as well as strongly coupled theories.
%\pagebreak

\vskip 10pt
\noindent
{\bf Acknowledgements} \\
We would like to thank S. Bhattacharyya for useful discussion. 
SM would like to thank ICTS-TIFR for hospitality 
under the associateship programme where a part of this work had been 
done. Work of PPP was supported in part by UGC-BRS fellowship, 
Government of India.      
\vskip 10pt
%%%%%%%%%%%%%%%%%%%%%%%%%%%%%%%%%%%%%%%%%%%%%%%%%%%%%%%%%
%\begin{thebibliography}{99}
\providecommand{\href}[2]{#2}
%\begingroup\raggedright

%%%%%%%%%%%%%%%%%%%%%%%%%%%%%%%%%%%%%%%%%%%%%%%%%%%%%%%%%%%%%%%%%%
%\end{enumerate}
%%%%%%%%%%%%%%%%%%%%%%%%%%%%%%%%%%%%%%%%%%%%%%%%%%%%%%%%%
\end{document}